\documentclass[twocolumn,superscriptaddress,longbibliography,aps,pra,floatfix]{revtex4-1}

\usepackage{graphicx}
\usepackage{bm}
\usepackage{color}
\usepackage{amsmath}
\usepackage{amssymb}
\usepackage{epstopdf}
\usepackage{gensymb}

\usepackage{enumitem}

\usepackage{float}

\usepackage{xcolor}

\usepackage[urlcolor=blue,colorlinks=true,citecolor=blue,linkcolor=blue,pdfstartview={FitH},bookmarks=false]{hyperref}

\graphicspath{{fig/}{./fig/}{.}}

\begin{document}

\title{Origin of monoclinic distortion and its impact on the electronic properties in KO$_2$}

\author{Olga~Sikora}
\email[corresponding author; e-mail: ]{sikora@wolf.ifj.edu.pl}
\affiliation{\mbox{Institute of Nuclear Physics, Polish Academy of Sciences, 
W. E. Radzikowskiego 152, PL-31342 Krak\'{o}w, Poland}}  

\author{Dorota~Gotfryd}
\email[e-mail: ]{dorota.gotfryd@fuw.edu.pl}
\affiliation{\mbox{Institute of Theoretical Physics, Faculty of Physics, 
University of Warsaw, Pasteura 5, PL-02093 Warsaw, Poland}}
\affiliation{\mbox{Institute of Theoretical Physics, Jagiellonian University, 
Profesora Stanis\l{}awa {\L}ojasiewicza 11, PL-30348 Krak\'ow, Poland}}

\author{Andrzej~Ptok}  
\email[e-mail: ]{aptok@mmj.pl}
\affiliation{\mbox{Institute of Nuclear Physics, Polish Academy of Sciences, 
W. E. Radzikowskiego 152, PL-31342 Krak\'{o}w, Poland}}

\author{Ma\l{}gorzata~Sternik}
\affiliation{\mbox{Institute of Nuclear Physics, Polish Academy of Sciences, 
W. E. Radzikowskiego 152, PL-31342 Krak\'{o}w, Poland}}

\author{Krzysztof~Wohlfeld}
\affiliation{\mbox{Institute of Theoretical Physics, Faculty of Physics, 
University of Warsaw, Pasteura 5, PL-02093 Warsaw, Poland}}

\author{Andrzej~M.~Ole\'s}
\affiliation{\mbox{Institute of Theoretical Physics, Jagiellonian University, 
Profesora Stanis\l{}awa {\L}ojasiewicza 11, PL-30348 Krak\'ow, Poland}}
\affiliation{Max Planck Institute for Solid State Research, Heisenbergstrasse 1, 
D-70569 Stuttgart, Germany}

\author{Przemys\l{}aw~Piekarz}
\email[e-mail: ]{piekarz@wolf.ifj.edu.pl}
\affiliation{\mbox{Institute of Nuclear Physics, Polish Academy of Sciences, 
W. E. Radzikowskiego 152, PL-31342 Krak\'{o}w, Poland}}

\date{\today}

\begin{abstract}
We use the density functional theory and lattice dynamics calculations to 
investigate the properties of potassium 
superoxide KO$_2$ in which spin, orbital, and lattice degrees of freedom 
are interrelated and determine the low-temperature phase. After calculating 
phonon dispersion relations in the high-temperature tetragonal $I4/mmm$
structure, we identify a soft phonon mode leading to the monoclinic $C2/c$ 
symmetry and optimize the crystal geometry resulting from this mode. 
Thus we reveal a displacive character of the structural transition with the 
group-subgroup relation between the tetragonal and monoclinic phases. We compare 
the electronic structure of KO$_2$ with antiferromagnetic spin order in the 
tetragonal and monoclinic phases. We emphasize that realistic treatment of 
the electronic structure requires including the local Coulomb interaction $U$ 
in the valence orbitals of the O$^-_2$ ions. The presence of the `Hubbard' 
$U$ leads to the gap opening at the Fermi energy in the tetragonal 
structure without orbital order but with weak spin-orbit interaction.  We 
remark that the gap opening in the tetragonal phase could also be obtained 
when the orbital order is initiated in the calculations with a realistic value 
of $U$. Finally, we show that the local Coulomb interactions and the finite 
lattice distortion, which together lead to the orbital order via the 
Jahn-Teller effect, are responsible for the enhanced insulating gap in the 
monoclinic structure.
\end{abstract}

\maketitle

\section{Introduction}
\label{sec.intro}

Despite of their simple stoichiometry, alkali $R$O$_2$ superoxides~\footnote{also called hyperoxides or dioxides}  
(with $R=$~Na, K, Rb, Cs) are complex magnetic systems with couplings between 
spin, orbital and lattice degrees of freedom. They consist of alkali cations 
$R$ and the superoxide O$_2^-$ anions, i.e., charged 
oxygen molecules with an additional electron. The superoxide ion is interesting
by itself and plays a role e.g. in biological processes~\cite{hayyan.hashim.16}. 
In the crystal structure it provides partially filled $p$-electron bands that 
are responsible for a rich variety of phenomena. 

Diffraction experiments~\cite{ziegler.meister.75,kanzig.labhart.76,kassatochkin.kotow.36,abrahams.kalnajs.55} 
revealed multiple structural phase transitions that can be related to orientation 
of oxygen molecules. At high temperature the crystal structure is of cubic NaCl 
type with disoriented oxygen molecules, and upon decreasing temperature the 
symmetry in KO$_2$, RbO$_2$, and CsO$_2$ is lowered to the body-centered tetragonal 
($I4/mmm$) with average O$_2$ orientation parallel to the tetragonal axis. 
With further cooling all three compounds transform to incommensurate phases 
and eventually reach low-temperature ordered structures with the onset of magnetic 
order~\cite{zumsteg.ziegler.74}.

Except for many similar properties, recent experimental and theoretical works show 
that even minor features distinguishing the alkali compounds, such as the extent of the oxygen molecule rotation from the tetragonal axis~\cite{astuti.miyajima.19} or the position of an alkali cation $p$-levels~\cite{riyadi.zhang.12}  can strongly influence their magnetic behavior. 
One of the reasons of this behavior is the competition between different exchange and 
superexchange mechanisms in systems which reveal significant geometrical and
spin-orbital frustration~\cite{solovyev.08,kim.kim.10, wohlfeld.daghofer.11}. 

\begin{figure*}[!t]
\includegraphics[width=\linewidth]{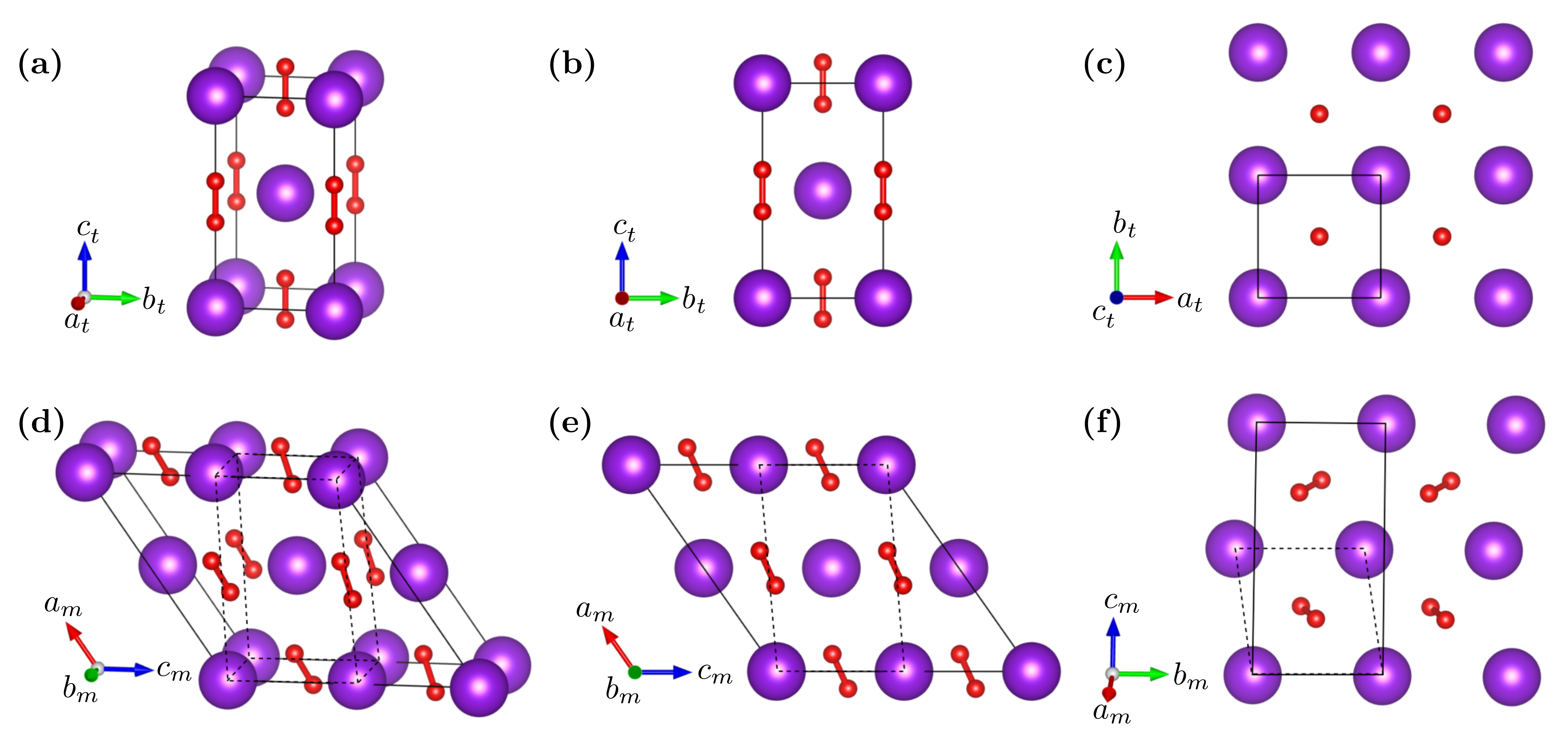}
\caption{
Schematic view of the crystal structures of the potassium dioxide KO$_{2}$.
Left panels show: (a) high temperature $I4/mmm$ tetragonal cell and 
(d) moderate temperature $C2/c$ monoclinic cell, with oxygen molecules 
tilted from the vertical direction. K and O$_2$ units are shown by 
violet balls and pairs of red balls. Panel (e) reveals the relation 
between the tetragonal cell shown along the $a_t$ direction in 
panel (b) with the monoclinic cell projected along the $b_m$ direction. 
Similarly, the relation between the bases of the cells for the 
tetragonal (c) and monoclinic (f) phase is shown for the planes  
$a_tb_t$ and $b_mc_m$, respectively. 
The~image was rendered using {\sc vesta} software~\cite{momma.izumi.11}.}
\label{fig.crys}
\end{figure*}

In potassium dioxide KO$_2$, the high temperature cubic structure~\cite{carter.margrave.52} 
transforms at $395$~K to the tetragonal $I4/mmm$ phase~\cite{kassatochkin.kotow.36,abrahams.kalnajs.55}, presented in Fig.~\ref{fig.crys}(a).
Below $231$~K, KO$_2$ is found in the incommensurate phase and at $196$~K its crystal 
structure stabilizes in the monoclinic $C2/c$ phase depicted in Fig.~\ref{fig.crys}(d).
Further lowering of the symmetry leads to the triclinic phase, where a long-range 
$C$-type antiferromagnetic (AFM) order, with opposite spin orientations in two adjacent 
ferromagnetic $(00l)$ oxygen layers was detected below the N\'eel 
temperature $T_{\rm N}\simeq 7$~K by the neutron scattering~\cite{smith.nicklow.66}. 
In early theoretical studies, the occurrence of the AFM order was connected to the 
coherent tilting of O$_2$ molecules observed in low-temperature phases~\cite{kanzig.labhart.76,labhart.raoux.79,lines.81}. The role of Jahn-Teller distortions 
at lower temperatures has been indicated~\cite{halverson.62}
and the effect of the O$_2$ tilting on the semiconducting propeties of KO$_2$ was discussed within the $p$-electron 
correlated models~\cite{khan.mahanti.75}.

The electronic band structure calculations for the tetragonal phase of potassium dioxide~\cite{solovyev.08,kim.kim.10}, performed within the density functional theory (DFT), 
found the metallic ground state in disagreement with the experimental observations. 
This indicates the importance of electron correlations,
and indeed, the combination of moderate $U$ and spin-orbit coupling (SOC) included within the 
GGA+$U$ approach opens an insulating gap of the order of $1.0$~eV~\cite{kim.kim.10,Kan10}. Larger values of the 
gap ($\sim$2-5 eV) were obtained within the hybrid functional calculations 
and non-self-consistent many-body perturbation theory performed for the tetragonal 
structure with the tilted O$_2^-$ molecules~\cite{mathiesen.yang.2019}.

The monoclinic structure of KO$_2$ expected to be more stable at low temperature has 
been studied using \textit{ab initio} methods and its formation energy compared to 
other potassium oxides~\cite{nandy.mahadevan.10,nandy.mahadevan.11,nandy.mahadevan.12}. 
The lattice distortion was found to be accompanied by two-sublattice orbital order~\cite{nandy.mahadevan.10}. In some other studies~\cite{kim.kim.10,kim.min.14}, 
a simplified structure was considered with potassium ions in the tetragonal positions and 
all oxygen molecules tilted coherently, what implies the ferro-orbital order. Due to the 
lowering of symmetry and the onset of orbital order, the phase with lattice distortion 
reveals an insulating gap in the presence of the interaction $U$ alone~\cite{kim.kim.10,kim.min.14}.

The studies presented until now focused mainly on the electronic properties of KO$_2$. 
They uncover the cooperative scenario, in which the local 
Coulomb interactions $\propto U$ and the Jahn-Teller effect play an important role 
in the low-temperature phase~\cite{halverson.62}. However, a detailed mechanism of 
the structural transition, which breaks the tetragonal symmetry and leads to the 
monoclinic $C2/c$ phase has not been explained so far. In particular, the lattice 
dynamics in KO$_2$ was never studied in the theory, and the role of phonons in the 
tetragonal-monoclinic transition is unknown.

Here we investigate the origin of the monoclinic phase and present a comprehensive 
description of its structural, dynamical, and electronic properties. We calculate 
phonon dispersion relations to demonstrate dynamical instability of the tetragonal 
cell at low temperatures, and use the lowest soft mode at the $N$ point of the 
Brillouin zone to generate and optimize the monoclinic $C2/c$ structure with 
rotated oxygen molecules and distorted potassium sublattice. We analyze its 
electronic properties and the origin of orbital order. The low-temperature 
magnetic order is also discussed. Special attention is given to possible mechanisms 
leading to the insulating behavior, i.e., the origin of the gap in the presence of 
either local Coulomb interaction $U$, or SOC, or both interactions acting jointly.  
 
The paper is organized as follows. 
The calculation method is described in Sec.~\ref{sec.com_details}.
In Sec.~\ref{sec.crys_stru} we discuss structural properties of the tetragonal 
and monoclinic phase of KO$_2$, and analyze the phonon dispersion relations. 
We show the soft mode in the tetragonal structure that leads to the monoclinic 
distortion and demonstrate the stability of the monoclinic phase. 
Sec.~\ref{sec.electron} presents electronic band structures in both phases,  
while we discuss the source of magnetic order in Sec.~\ref{sec.magnetic}, and 
the origin of orbital order in the low temperature structure in 
Sec.~\ref{sec.orbital}. Electronic densities of states are investigated in 
Sec.~\ref{sec.gap}. To uncover the mechanisms responsible for the gap opening, 
we consider models with either the Coulomb $U$, or with the SOC or both. 
In Sec.~\ref{sec.summary} we conclude with a summary of the results and 
discussion of open questions.
Appendix~\ref{sec.appA} presents calculations for the tetragonal cell in 
which an orbital order is initiated and stabilized by the Coulomb $U$.
Finally, Appendix~\ref{sec.appB} shows the electronic densities of states 
with partial distortions of the tetragonal structure, to show separately the 
impact of the potassium and oxygen displacements on the electronic structure.

%%%%%%%%%%%%%%%%%%%%%%%%%%%%%%%%%%%%%%%%%%%%%%%%%%%%%%%%%%%%%%%%
\section{Computational details}
\label{sec.com_details}

The spin-polarized DFT calculations were performed within the projector augmented-wave (PAW) method~\cite{blochl.94} using the Vienna 
\textit{Ab initio} Simulation Package ({\sc vasp})~\cite{kresse.hafner.94,kresse.furthmuller.96,kresse.joubert.99}.
The exchange-correlation potential was obtained by the generalized 
gradient approximation (GGA) in the form proposed by Perdew, Burke, and Enzerhof (PBE)~\cite{pardew.burke.96}. 
Strong local electron interactions were 
included within the DFT+$U$ scheme~\cite{liechtenstein.anisimov.95}, and we 
assume the intraorbital Coulomb parameter $U=4.0$~eV as in the earlier studies~\cite{nandy.mahadevan.10,riyadi.zhang.12}, and Hund's exchange $J=0.62$~eV~\cite{solovyev.08} on the oxygen $2p$ orbitals. We also investigated the impact 
of the SOC~\cite{steiner.khmelevskyi.16} on the electronic structure.

We performed calculations for the AFM spin order found in the experiment~\cite{smith.nicklow.66}, taking into account noncollinear spin configurations 
in calculations for the monoclinic phase. Optimization of both the structural 
parameters and the electronic structure in the tetragonal (monoclinic) phase 
were performed using a $16\times 16\times 8$ ($8\times 16\times 8$) 
Monkhorst-Pack grid~\cite{monkhorst.pack.76} of {\bf k}-points. 
The energy cut-off for the plane-wave expansion was equal to 520 eV.
We checked the convergence of the structural parameters for the tetragonal phase by increasing its value to 700 eV and found very small changes (by less than $0.3\%$) of the lattice constants and the O--O bond length. The structures were relaxed using the conjugate gradient technique with the 
energy convergence criteria set at $10^{-7}$~and $10^{-5}$~eV for the 
electronic and ionic iterations, respectively. The electronic density 
of states (DOS) was calculated using the tetrahedron method.

For the relaxed structures, the phonon dispersion relations as well as the total 
and element-projected phonon DOS were calculated using the direct 
method~\cite{parlinski.li.97} implemented in the {\sc Phonon} software~\cite{phonon}. 
In this approach, the Hellmann--Feynman forces acting on all atoms in a given 
supercell are obtained by single-atom displacements from their equilibrium positions.
We note that such atomic displacements do not change the insulating state of the system.
The force-constant matrix elements are derived using the singular-value 
decomposition (SVD) technique and then the dynamical matrix is obtained. To describe 
the longitudinal optic--transverse optic (LO--TO) splitting induced by macroscopic 
polarization, the static dielectric tensor and Born effective charges were 
determined using density functional perturbation theory~\cite{gajdos.hummer.06}. 
The phonon energies and polarization vectors are calculated by the exact 
diagonalization of the dynamical matrix.

\section{Crystal structures and their stability}
\label{sec.crys_stru}

\subsection{Tetragonal phase}

By relaxing the tetragonal cell with two KO$_2$ formula units, see 
Fig.~\ref{fig.crys}(a), we obtained the lattice constants \mbox{$a_{t}=3.979$~\AA} 
and $c_{t} = 6.928$~\AA, both in a good agreement with the experimental data, 
$a_{t}=4.034$~\AA\ and $c_{t}=6.699$~\AA~\cite{kassatochkin.kotow.36,abrahams.kalnajs.55}.
In this structure, oxygen molecules are oriented along the $c_{t}$ direction. 
The calculated O--O bond length of $1.361$~\AA\ is larger than the experimental values of 1.28(7)~\AA~\cite{kassatochkin.kotow.36} and 1.28(2)~\AA~\cite{abrahams.kalnajs.55} typically presented in the literature.  
In fact, these values are underestimated because the librational motions of O$_2$ molecule are not considered in structural models used to evaluate the diffraction data~\cite{hesse.jansen.89,dietzel.kremer.04}. 
Reinterpretation of the same data with a model allowing the motion of the O$_2$ molecule around the $c$ axis leads to a bond length in the range of 1.32-1.35~\AA~\cite{halverson.62}. 
Similar values, ranging between 1.31 and 1.34~\AA, are found in the study performed using the near edge X-ray absorption fine structure experiment and multiple scattering calculations on KO$_2$~\cite{pedio.wu.02}, and in other studies on various superoxides~\cite{seyeda.jansen.98,dietzel.kremer.04}.

\begin{figure*}[!pt]
\includegraphics[width=0.95\linewidth]{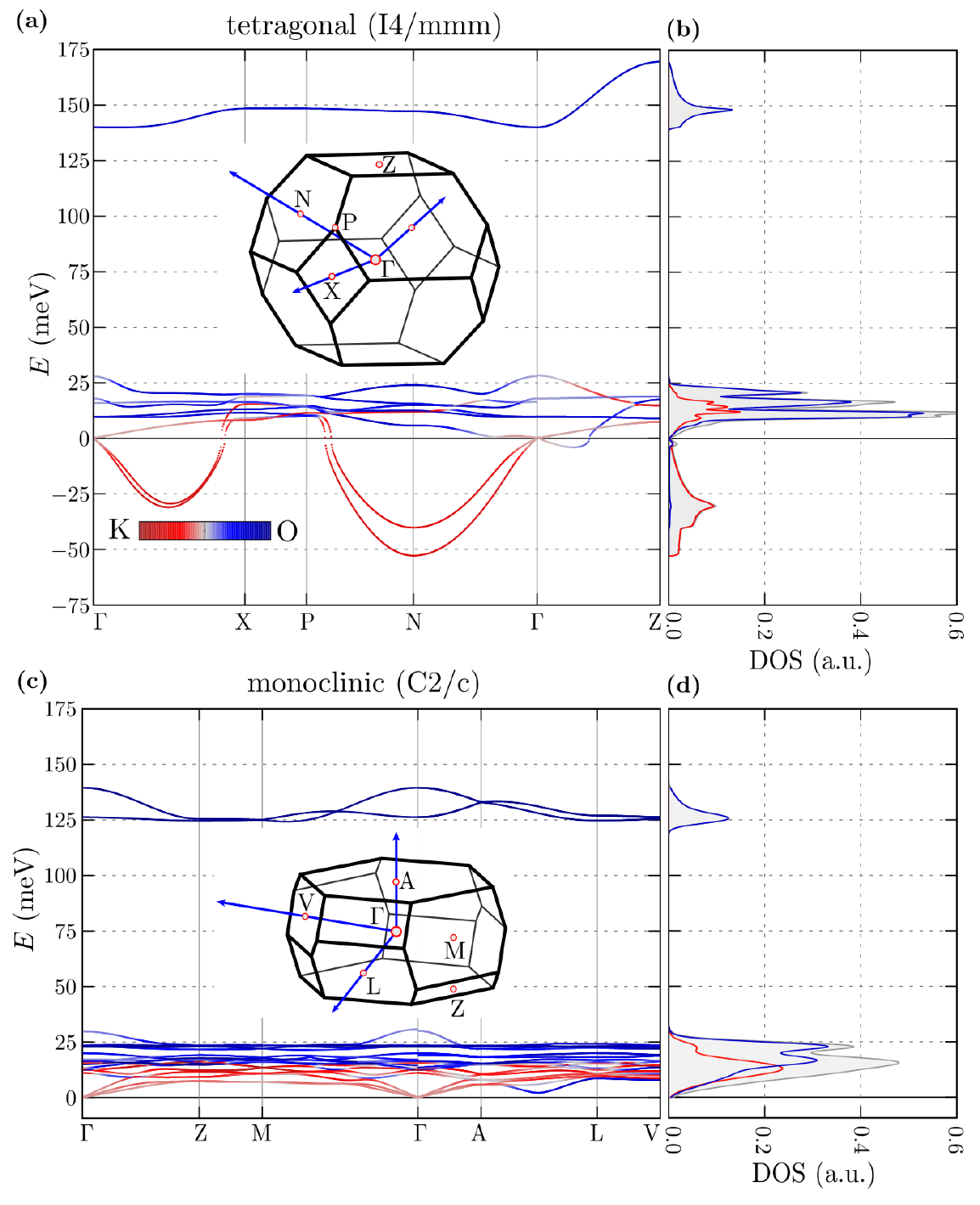}
\caption{Phonon dispersion curves along high symmetry directions
[cf. the first Brillouin zone (FBZ) in the insets]  for: 
(a) the tetragonal $I4/mmm$ and (c) monoclinic $C2/c$ phase of
KO$_2$. The color of lines 
indicates the contribution of K or O atoms (cf. color bar). In the right panels, the phonon density of states (DOS) is shown for (b) tetragonal and (d) monoclinic phase, with gray, red and blue corresponding to the total, partial K and partial O phonon DOS, respectively.
The lattice dynamics calculations of the tetragonal structure (a) reveal several 
soft modes with imaginary frequency (presented as negative values), predominantly 
associated with K atoms. The lowest mode at the N point leads to the monoclinic 
symmetry ($C2/c$). Phonon dispersion relations for the optimized 
$C2/c$ structure (c) do not reveal the soft modes, confirming the stability of 
the monoclinic phase.}
\label{fig.ph_all}
\end{figure*}

The phonon dispersion relations calculated for the tetragonal (insulating) structure are presented in Fig.~\ref{fig.ph_all}(a), along the high symmetry directions of the first 
Brillouin zone (FBZ) shown in the inset, and the phonon DOS is displayed in 
Fig.~\ref{fig.ph_all}(b). The contributions of oxygen and potassium atoms to 
individual phonon branches and DOS are depicted in blue and red, 
respectively. The entire phonon spectrum of KO$_2$ consists of two separated 
energy ranges. The low-energy part below $30$~meV includes the acoustic and 
optic vibrations of K and O atoms.  In contrast, the high-energy dispersion branch
around $150$~meV corresponds to the internal vibrations in the individual oxygen molecules. 
Two phonon branches exhibit imaginary frequencies (plotted with negative values)
along some directions of the reciprocal 
space, reflecting dynamical instability of the tetragonal structure at low 
temperatures. These imaginary modes arise predominantly due to the vibrations 
of K atoms, however, small contributions of the oxygen movements are also visible 
in the partial phonon DOS presented in Fig.~\ref{fig.ph_all}(b).

At the $\Gamma$ point, the group theory predicts the following decomposition 
of the irreducible representations of the phonon modes: 
$\Gamma=A_{1g}+2A_{2u}+E_g^{(2)}+2E_u^{(2)}$, where $A_{1g}$ and $E_g$ are 
Raman-active modes, while $A_{2u}$ and $E_u$ are infrared-active modes. 
In agreement with the dimension of irreducible representations, all $E$ 
modes are doubly degenerate at the FBZ center. The energies of the TO 
modes at the $\Gamma$ point are collected in Tab.~\ref{tab_gamma}. Two 
infrared optical modes exhibit the LO--TO splitting and the LO mode is 
shifted to higher energies due to the long-range electric polarization. 
The Raman measurements of KO$_2$ performed at $300$~K confirm the presence 
of the strong scattering at $141.48$~meV~\cite{bates.1972}, which is 
attributed to the stretching mode of O$_2^{-}$ ions. 

After examining the crystal structure resulting from atomic displacements 
induced by the polarization vector of the lowest soft mode at the N point, we 
found the $C2/c$ monoclinic symmetry. 
This soft mode 
rotates oxygen molecules and induces shifts of potassium atoms along the $b_t$ direction. 
It shows that the tetragonal-to-monoclinic 
transition has a displacive character with the group-subgroup relation between 
the high-temperature and low-temperature symmetries. The group theory analysis 
indicates a few possible intermediate symmetries (subgroups of $I4/mmm$ and 
supergroups of $C2/c$), which may explain the existence of the incommensurate 
phase observed by diffraction studies~\cite{ziegler.meister.75}. 

\subsection{Monoclinic phase}

We have obtained the monoclinic $C2/c$ structure using the soft mode revealed by 
the calculations for the tetragonal structure, as discussed above. The monoclinic 
structure has a lower total energy with respect to the tetragonal phase, 
approximately $0.129$~eV per one formula unit (f.u.).
For both phases, the total energies were calculated for the insulating AFM states.
\begin{table}[!t] 
\caption{Phonon modes with their irreducible representations (IR), energies, 
and activities (R--Raman and I--infrared), as obtained from the phonon calculations 
at the $\Gamma$ point in the tetragonal and monoclinic phases of the potassium 
dioxide KO$_2$. For the infrared phonons only the TO modes are indicated. 
In parentheses the measured energies of Raman modes~\cite{bates.1972} are 
presented (when available). }
\begin{ruledtabular}
\begin{tabular}{crrc|crrc}
\multicolumn{4}{c}{Tetragonal $I4/mmm$} & \multicolumn{4}{c}{Monoclinic $C2/c$} \\
 IR & \multicolumn{2}{c}{$E$ (meV)} & Activity & IR & \multicolumn{2}{c}{$E$ (meV)} & Activity \\
\hline
$E_g$    &   9.73 &          &  R &  $B_g$ & 11.09 & (8.06)  &  R \\
$A_{2u}$ &  15.98 &          &  I &  $A_u$ & 12.55 &         &  I \\
$E_u$    &  18.00 &          &  I &  $A_g$ & 12.59 & (10.04) &  R \\
$A_{1g}$ & 139.99 & (141.48) &  R &  $B_g$ & 13.35 & (11.04) &  R \\
         &      &            &    &  $B_u$ & 13.51 &         &  I \\
         &      &            &    &  $B_g$ & 15.65 & (15.13) &  R \\
         &      &            &    &  $A_g$ & 16.38 & (17.36) &  R \\
         &      &            &    &  $B_u$ & 17.01 &         &  I \\
         &      &            &    &  $A_u$ & 19.91 &         &  I \\
         &      &            &    &  $B_u$ & 21.21 &         &  I \\
         &      &            &    &  $A_g$ & 23.00 & (25.42) &  R \\
         &      &            &    &  $B_g$ & 23.46 & (26.16) &  R \\
         &      &            &    &  $A_u$ & 23.47 &         &  I \\
         &      &            &    & $B_g$ & 126.21 &         &  R \\
         &      &            &    & $A_g$ & 139.47 & (141.98) & R 
\end{tabular}
\end{ruledtabular}
\label{tab_gamma}
\end{table}
Within the optimization procedure, we found the lattice parameters, $a_{m}=8.098$~\AA, 
$b_{m}=4.200$~\AA, \mbox{$c_{m}=8.178$~\AA,} and $\beta=125.093\degree$, similar
to those obtained from the x-ray diffraction: $a_{m} =7.880$~\AA, $b_{m} = 4.036$~\AA, 
$c_{m} =7.968$~\AA, and $\beta=122.85\degree$~\cite{ziegler.meister.75}. 
The calculated O--O bond length is 1.359~\AA, only slightly smaller than in the tetragonal phase.
In the monoclinic cell two nonequivalent atoms K and O are placed at the Wyckoff positions 
$4e(0,0.078,0.25)$ and $8f(0.416,0,072,-0.017)$, respectively. We note that in 
contrary to the analysis of the experimental data~\cite{ziegler.meister.75} 
and some theoretical studies~\cite{kim.kim.10,kim.min.14}, 
we find pronounced changes in the potassium positions with respect to the tetragonal 
structure. The relation between both crystal structures is shown in Fig.~\ref{fig.crys}.
The lattice parameters in the tetragonal and monoclinic phases are connected by the 
following approximate relations: $a_m \sim \sqrt{a_t^2+c_t^2}$, $b_m \sim b_t$, 
and $c_m \sim 2a_t$ [cf.~Figs.~\ref{fig.crys}(a) and~\ref{fig.crys}(d)]. The number 
of atoms in the monoclinic unit cell is twice as large as in the tetragonal one.

The lattice deformation induced by the structural transformation is best 
visible in the projections presented in Fig.~\ref{fig.crys}. The relation 
between positions of K atoms in the bases of the tetragonal and monoclinic 
cell is presented in Fig.~\ref{fig.crys}(f).
The shifts of K atoms are harmonized with the rotations of O$_2^-$ anions.
These rotations can be described as the coherent tilting ($24.4\degree$)
in the $a_mc_m$ plane shown in Fig.~\ref{fig.crys}(e), combined with rotations  
alternating along the $c_m$ direction [cf.~Fig.~\ref{fig.crys}(f)]. The total 
tilting of the molecule from the tetragonal $c_t$ direction amounts to $35.3\degree$.  
Similar deformation was obtained by the relaxation of the KO$_2$ structure 
without the symmetry constraints in Ref.~\cite{nandy.mahadevan.10}.
The formation energy of KO$_2$ was calculated for the $C2/c$ symmetry, however, 
a different setting of the lattice parameters for the distorted structure was 
chosen~\cite{nandy.mahadevan.11,nandy.mahadevan.12}.
We note that a similar $C2/c$ structure was predicted in SrO$_2$ from 
first-principles swarm structure searching simulations~\cite{wang.wang.17}.

\begin{figure}[!b]
\begin{center}
\includegraphics[width=\linewidth]{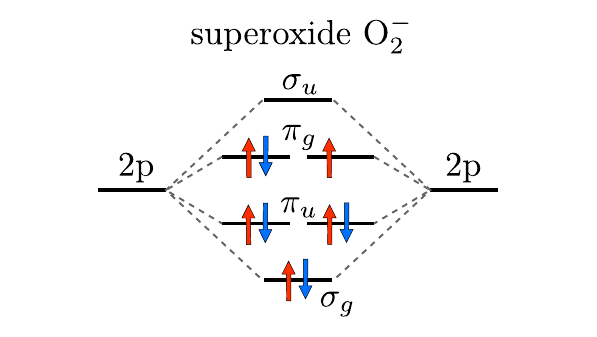}
\end{center}
\caption{Electronic energy levels for an {\it isolated} O$_2$ molecule 
formed by the oxygen $2p$ orbitals of both oxygen atoms, and their occupancy 
in the superoxide anion O$_2^-$ filled by nine electrons.
}
\label{fig.ko2_mag}
\end{figure}

\begin{figure*}[t!] 
\begin{center}
\includegraphics[width=\linewidth]{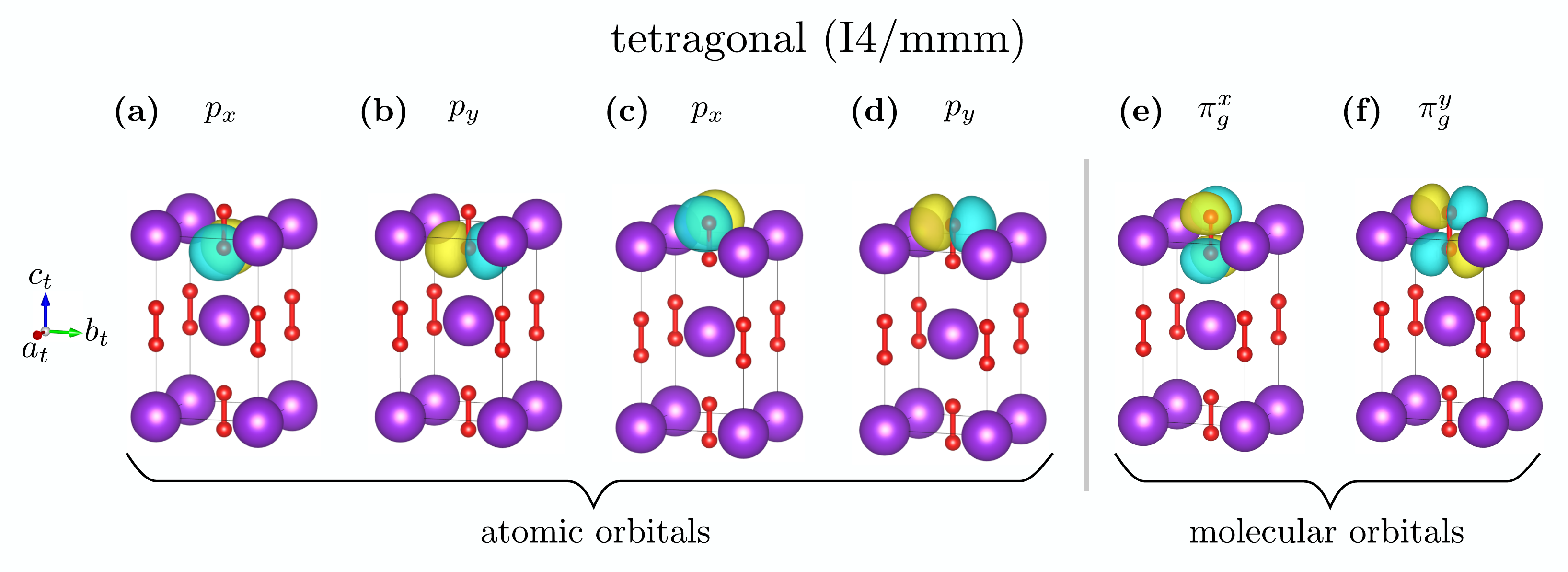}
\end{center}
\caption{
Four atomic and two molecular valence orbitals in the tetragonal phase: 
(a)-(d) $p_x$ and $p_y$ atomic orbitals of the  oxygen 
atoms,
(e)-(f) $\pi^x_{g}$ and $\pi^y_{g}$ molecular orbitals of an O$_2^-$ molecule. 
The presented orbitals were obtained from the numerical analyzes of the DFT 
bands around the Fermi level.
The image was rendered using {\sc vesta} software~\cite{momma.izumi.11}.
}
\label{fig.w90_orb}
\end{figure*}

The monoclinic $C2/c$ structure is dynamically stable as shown in Fig.~\ref{fig.ph_all} 
(lower panels). Twice as many atoms in the primitive unit cell 
generate the doubled number of phonon branches in the monoclinic structure 
compared to tetragonal one, and because of low symmetry all degenerate modes are 
split. For both structures the upper limit of the low-energy part is very similar, 
but in the  monoclinic phase two highest dispersion branches are shifted to lower 
energies. All phonons at the $\Gamma$ point are described by one-dimensional 
irreducible representations: $\Gamma=4A_g+4A_u+5B_g+5B_u$, where $A_g$, $B_g$ 
are Raman-active modes and $A_u$, $B_u$ are infrared-active modes. Their 
energies are presented in Tab.~\ref{tab_gamma}. For the Raman modes, we obtained 
a good agreement with the experimental values measured at $80$~K~\cite{bates.1972}.  

%%%%%%%%%%%%%%%%%%%%%%%%%%%%%%%%%%%%%%%%%%%%%%%%%%%%%%%%%%%%%%%%%%

\section{Electronic properties} 

\subsection{Molecular orbitals for an $O_2^-$ molecule}
\label{sec.O2-}

The KO$_2$ crystal is ionic and thus its electronic properties are 
strongly associated with the orbital structure of the O$_2^-$ ions, 
in which the hybridization between atomic $2p$ orbitals gives rise to bonding 
$\{\sigma_{g},\pi_{u}\}$ and antibonding $\{\pi_{g},\sigma_{u}\}$ molecular 
orbitals, see Fig.~\ref{fig.ko2_mag}. An {\it isolated} O$_{2}$ molecule with eight 
$2p$ electrons has two unpaired electrons in the doubly degenerate $\pi_g$ level 
and a triplet ($S=1$) ground state, as required by the first Hund's rule. 
KO$_2$ however includes O$_2^-$ anions with one extra electron 
coming from the alkali ion and the molecular configuration is~\cite{khan.mahanti.75}  
$\sigma_{g}^{2} \pi_{u}^{4} \pi_{g}^{3}$, as shown in Fig.~\ref{fig.ko2_mag}. 

\begin{figure}[b!]
\includegraphics[width=1.04\linewidth]{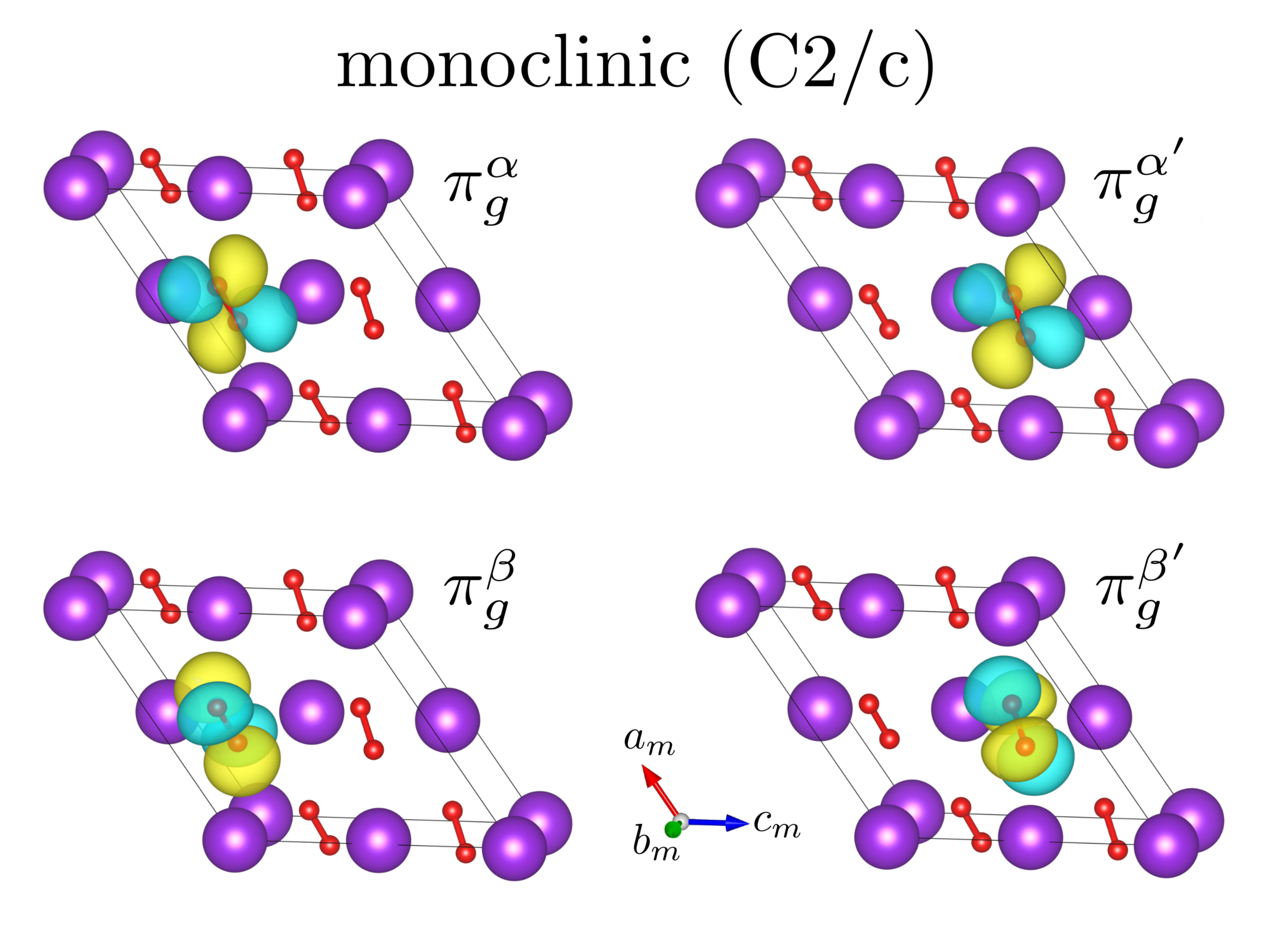}
\caption{
Two inequivalent pairs of molecular valence orbitals in the monoclinic phase: 
\{$\pi^\alpha_{g}$, $\pi^\beta_{g}$\}
and  
\{$\pi^{\alpha'}_{g}$, $\pi^{\beta'}_{g}$\},  corresponding to two orientations of the oxygen molecule. 
The presented orbitals were obtained from the numerical analysis of the 
DFT bands around the Fermi level.
The image was rendered using {\sc vesta} software~\cite{momma.izumi.11}.
}
\label{fig.w90_orb_c2c}
\end{figure}

\begin{figure*}[t]
\includegraphics[width=1.04\linewidth]{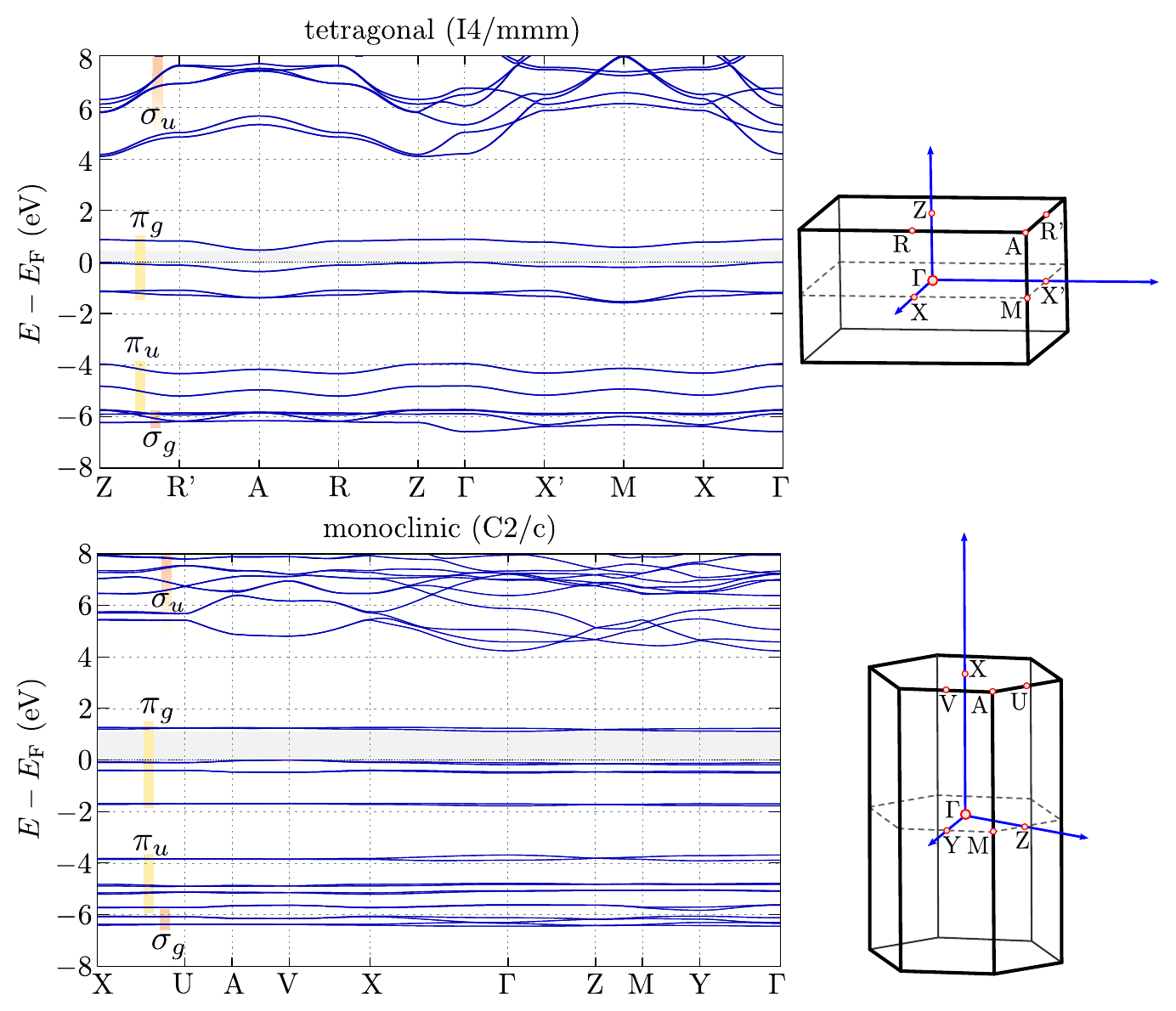}
\caption{
Electronic band structure for the tetragonal (top panels) and monoclinic (bottom panels) 
phases in the magnetic unit cell. Right panels present the first (magnetic) Brillouin 
zones. Results for GGA+$U$+SOC calculations in the presence of the $A$-AFM order 
(cf. Fig.~\ref{fig.mom_mag}).
}
\label{fig.el_bands_muc}
\end{figure*}

Important differences between the tetragonal and monoclinic phases, that may influence
their electronic properties, are already visible when the molecular orbitals of the
O$_2^-$ ion are considered either in the tetragonal or monoclinic crystal lattice.
In the tetragonal phase, the molecular $\pi_g$ orbitals 
[Fig.~\ref{fig.w90_orb}(e)-(f)], located around the Fermi level, are combinations of 
either two $p_x$ or two $p_y$ atomic orbitals 
[cf. Fig.~\ref{fig.w90_orb}(a)-(d)].
We note that different combinations of these atomic orbitals lead to the fully occupied
$\pi_u$ states, and the occupied bonding $\sigma$ orbital is 
composed of two $p_{z}$ orbitals (not shown). 
Transition from the tetragonal to 
monoclinic phase is associated with tilting of the O$_2^-$ ions, and this 
also leads to rotation of the molecular orbitals in space, as shown for the 
$\pi_{g}$-type orbitals in Fig.~\ref{fig.w90_orb_c2c}.
Because of two inequivalent rotations of the oxygen ions, two different sets of orbitals
are needed as bases for the valence $\pi_g$ states: $\{\alpha,\beta\}$ on one sublattice
and $\{\alpha',\beta'\}$ on the other.

\subsection{Electronic band structure}
\label{sec.electron}

The electronic band structure (Fig.~\ref{fig.el_bands_muc}) shows a few 
characteristic features that do not depend on the crystal geometry and the calculation 
scheme (i.e., including spin-orbit coupling or local Coulomb interaction).
The $\sigma_{g}$ bands are located around $-6$~eV below the Fermi level.
Slightly higher the bands formed by the bonding $\pi_{u}$ molecular orbitals
are located, while a group of the antibonding $\pi_{g}$ bands is situated around 
the Fermi level. In all these bands the role of the potassium orbitals is 
negligible~\cite{nandy.mahadevan.10}. The antibonding  $\sigma_{u}$ orbitals are 
strongly intermixed with empty K states above $4$~eV.

In the tetragonal phase, the splitting of the $\pi_{g}$ bands at the $\Gamma$ point 
is equal to approximately $25$~meV, what is similar to the value reported 
previously in Ref.~\cite{solovyev.08}. Contrary to dioxides of $4d$ or $5d$ 
elements, i.e., RuO$_2$~\cite{jovic.koch.18} or IrO$_2$~\cite{khan.poll.14,sun.zhang.17,das.slawinska.18,xu.jiang.19}, the value of the 
SOC is relatively small in KO$_2$. In fact, the SOC value strongly depends on 
the mass of compound components and plays an important role for massive
atoms~\cite{herman.kuglin.63,shanavas.popovic.14}, while here both atoms forming the 
structure are relatively light. In the monoclinic phase, the value of SOC can 
decrease by tilting of the O$_2$ molecules~\cite{solovyev.08}. 

The most apparent difference between the tetragonal and monoclinic electronic 
band structure is the increase of the charge gap from $0.5$~eV in the tetragonal 
phase to $1.1$~eV in the monoclinic phase, which is in a relatively good agreement 
with the activation energy $1.3$~eV obtained from the electrical conductivity 
measurements~\cite{khan.mahanti.75}. Also from x-ray absorption spectroscopy 
(XAS) measurements, a similar value of the gap was found~\cite{Kan10}. This 
increase can be associated with the onset of the orbital order stabilized by the 
lattice distortion (discussed in more detail in Sec.~\ref{sec.gap} and the Appendices). 
Due to the folding of the Brillouin zone, we observe that the band degeneracy 
doubles. Additionally, we observe ``flattening'' of the bands in the monoclinic 
phase in relation to the tetragonal phase. It corresponds well with the experimental 
results, which indicate that the main effect of the structural phase transition 
on the electronic structure is the band narrowing~\cite{khan.mahanti.75}.

\subsection{Magnetic order} 
\label{sec.magnetic}

\begin{figure}[b!]
\includegraphics[width=\linewidth]{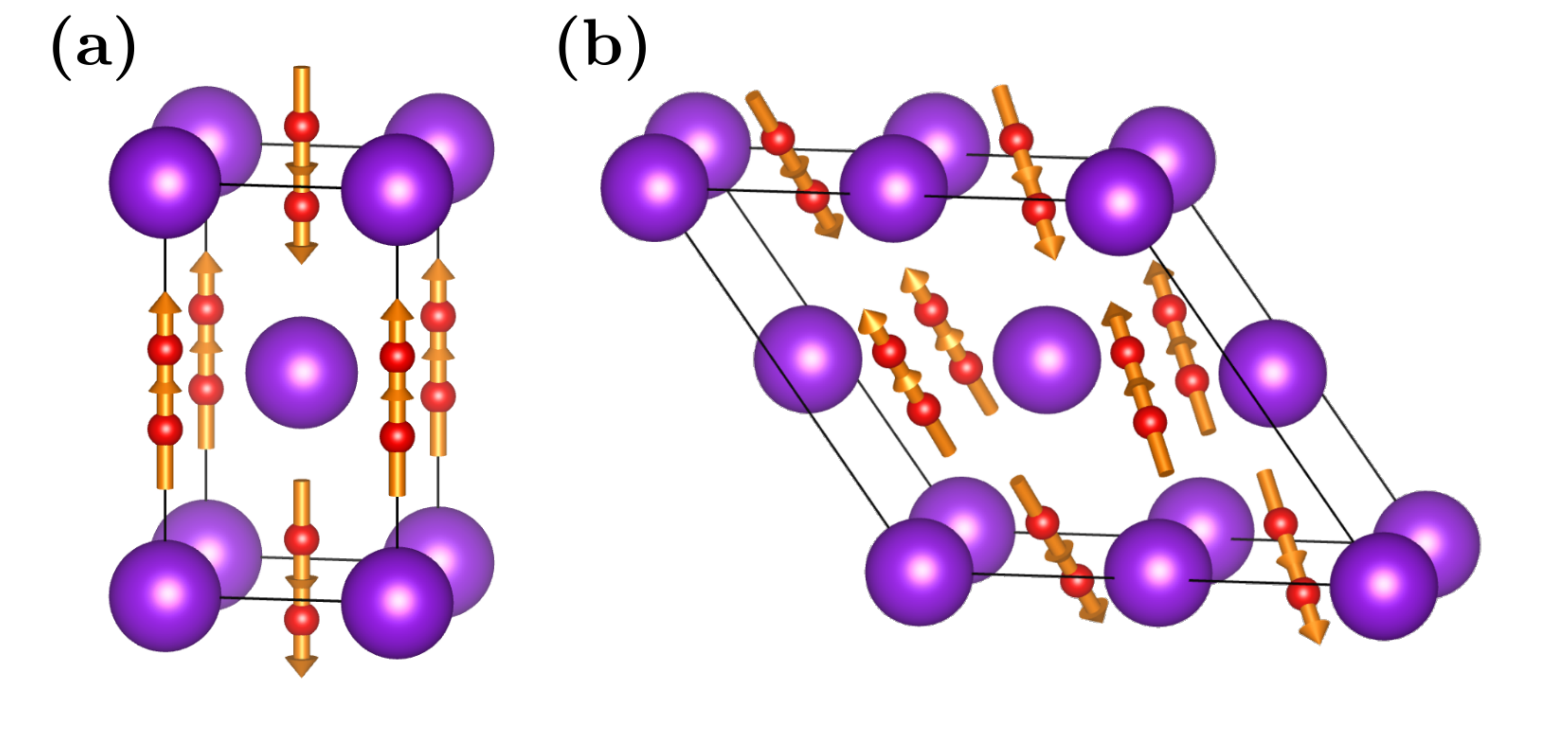}
\caption{The schematic view of the assumed 
$A$-AFM order in KO$_2$:
(a)~tetragonal and (b) monoclinic phase. Consecutive FM planes
along the $c_t$ axis have alternating magnetic moments on O$_2^-$ molecules;
conventions are the same as in Fig.~\ref{fig.crys}.
\mbox{The image was} rendered using {\sc vesta} software~\cite{momma.izumi.11}.
} 
\label{fig.mom_mag}
\end{figure}

All results presented in this work are obtained for the $A$-type AFM ($A$-AFM) order~\footnote{The $A$-type AFM ordering means ferromagnetic planes which are stacked antiferromagnetically, cf. the $A$-AFM order discussed in the context of e.g.: LaMnO$_3$~\cite{Wollan1955}, GdFeO$_3$~\cite{Zhu2017} or the several double perovskites~\cite{Aharbil2016}. Note that in Ref.~\cite{wohlfeld.daghofer.11} the same type of magnetic order as discussed here was referred to as $C$-type AFM, for in that paper the types of magnetic order were defined by the ratio of the ferromagnetic to the antiferromagnetic bonds.},
see Fig.~\ref{fig.mom_mag}. 
The investigated spin configuration for the tetragonal cell is  presented in Fig. 
\ref{fig.mom_mag}(a), with spins parallel to the oxygen molecule (we found the lowest 
energy for this spin configuration, although the values obtained for other orientations  
are very close). 
In the monoclinic phase, we assume spin directions to follow the rotation of molecules, 
therefore, the spin configuration is non-collinear as shown in Fig.~\ref{fig.mom_mag}(b).
Nevertheless, the projections of the magnetic moments on the $c_t$ axis maintain the 
$A$-AFM order. 

Earlier electronic structure calculations within the GGA+$U$ method showed lower energy 
of the ferromagnetic (FM) order with respect to the AFM state for small values of $U$ 
at oxygen ions, and the decreasing energy difference between these phases with growing 
$U$~\cite{nandy.mahadevan.10}. 
With the increasing value of $U$, the character of electronic states changes
from itinerant, which prefers the FM order, to localized, in which the FM and AFM
interactions compete due to the anisotropic kinetic exchange
between the O$_2^-$ anions~\cite{kim.kim.10}. 
The studies performed for a realistic spin-orbital model~\cite{wohlfeld.daghofer.11}
demonstrated that the geometric frustration of magnetic interactions can be lifted and the
$A$-AFM state is stabilized due to the orbital ordering.

Our calculations for the relaxed $C2/c$ structure show indeed very small energy 
differences between the FM and AFM configurations with slightly lower energy for the 
FM state regardless of the value of $U$. The negligibly small value $0.375$~meV/f.u. 
for $U=4$~eV confirms that magnetic interactions in KO$_2$ are highly frustrated~\cite{wohlfeld.daghofer.11}. Including additional effects like the superexchange 
and spin quantum fluctuations, not captured properly in the DFT+$U$ methods, may 
lower the energy of the $A$-AFM state. 

\subsection{Orbital order} 
\label{sec.orbital}

\begin{figure}[b!]
\includegraphics[width=\linewidth]{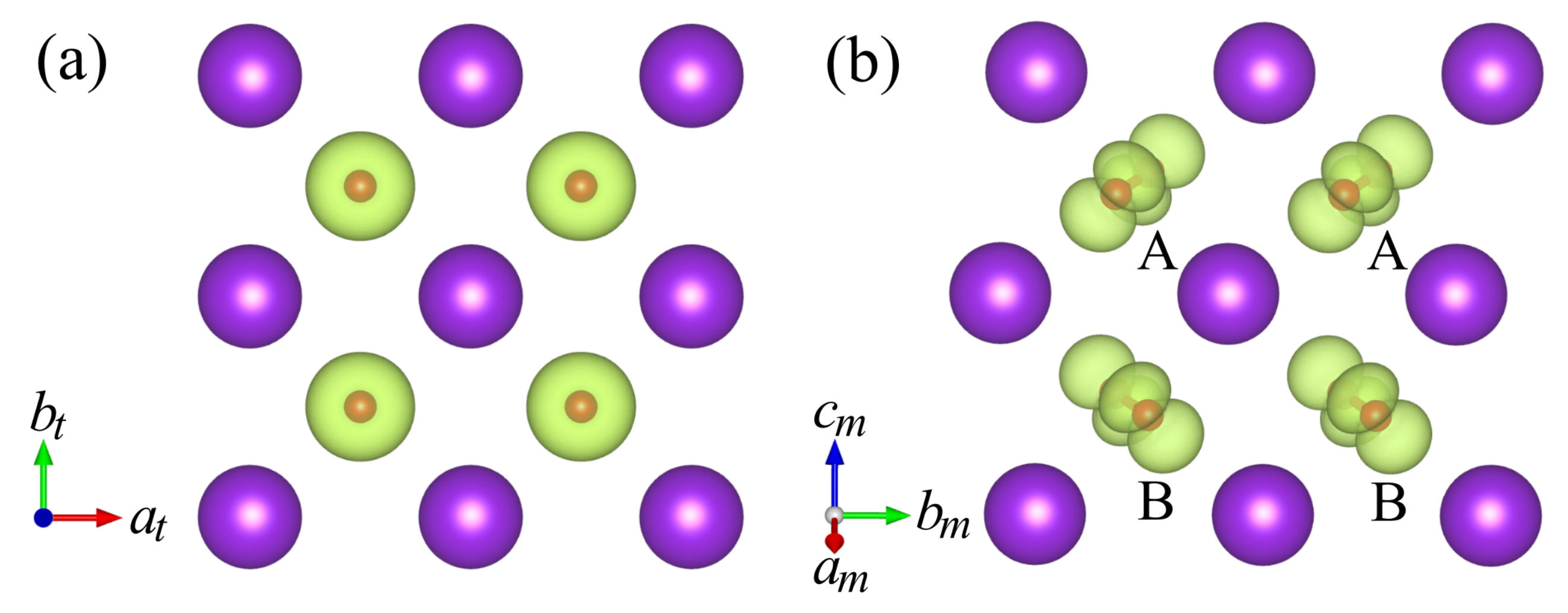}
\caption{
Calculated isosurfaces of the hole density (in green) for the: 
(a) tetragonal and 
(b) monoclinic phase. 
In the tetragonal phase, due to the degeneracy of the oxygen ion orbitals 
$\pi_g^x$ and $\pi_g^y$, the density distribution is symmetric around the 
molecule axis. In the monoclinic phase the orbital order emerges as a result 
of the lattice distortion. The unoccupied orbitals are linear combinations 
of the molecular orbitals $\{\pi_g^{\alpha'},\pi_g^{\beta'}\}$ and 
$\{\pi_g^\alpha,\pi_g^\beta\}$ on the sublattice A and B, respectively 
(cf.~Fig.~\ref{fig.w90_orb_c2c}).
The image was rendered using {\sc vesta} software~\cite{momma.izumi.11}.
}
\label{fig.c2_oo}
\end{figure}

\begin{figure*}[t!]
\includegraphics[width=\linewidth]{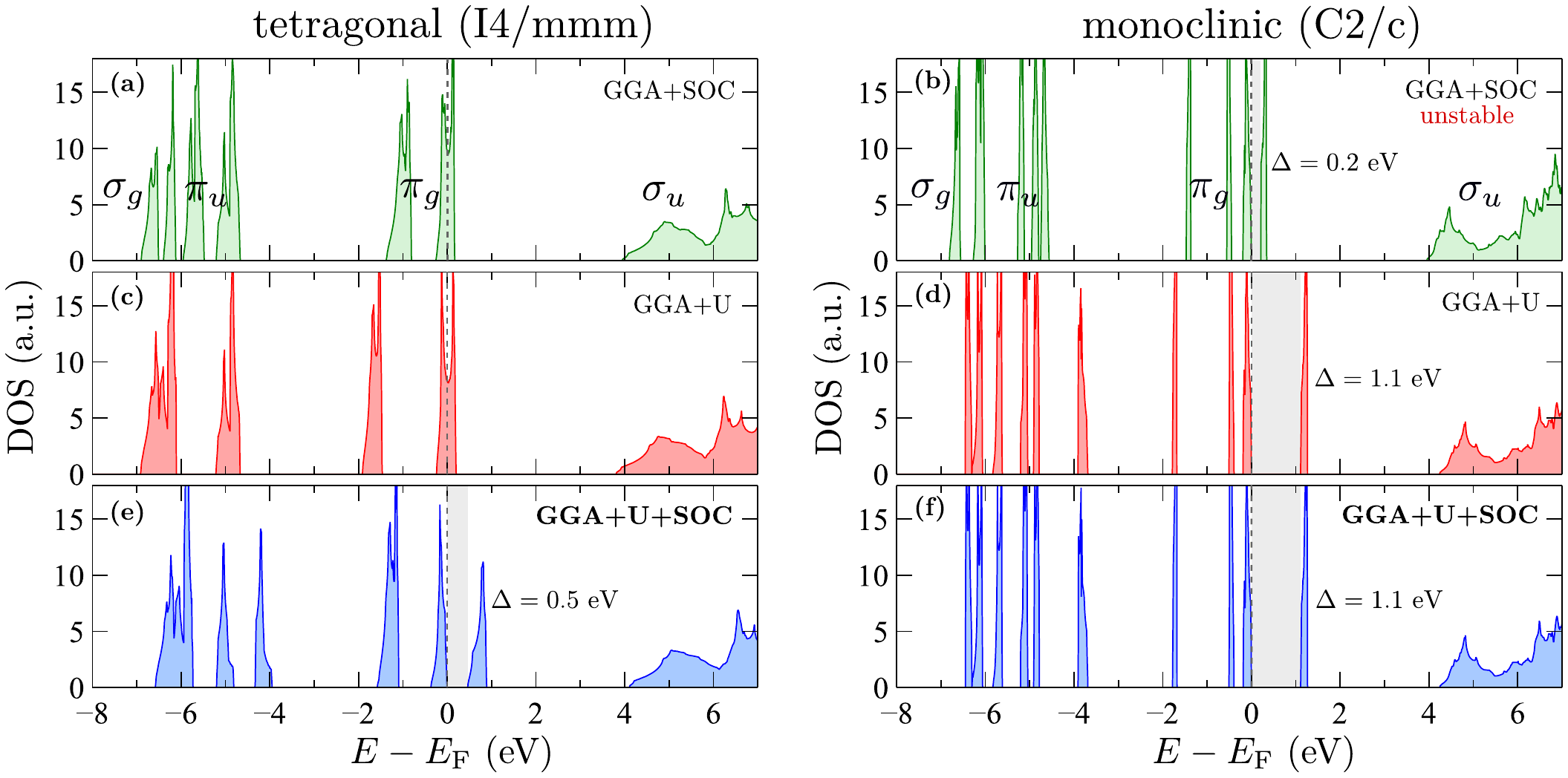}
\caption{
Comparison of the electronic density of states (DOS) obtained for the tetragonal 
and the monoclinic structures (left~and~right panels, respectively) within GGA+SOC, 
GGA+$U$, and GGA+$U$+SOC calculations (panels from top to bottom).
The assignment of each band to a particular O$_2$ molecular orbital character is 
indicated in the top panels. The insulating gap $\Delta$ is defined as the difference 
between the energies of the last occupied and first unoccupied states.
}
\label{fig.edos}
\end{figure*}

Calculations for the tetragonal cell performed with full symmetry do not break 
the degeneracy between the $\pi_g^x$ and $\pi_g^y$ orbitals. Their equal 
occupation can be demonstrated by plotting isosurfaces of the electronic density 
[see Fig.~\ref{fig.c2_oo}(a)] for the highest in energy, empty $\pi_g$ band 
(i.e., the hole density) in the $a_{t}b_{t}$ plane. Such calculations are 
analogous to restricted Hartree-Fock approach, while it is well known that 
a minimal energy is obtained only when the symmetry is broken in the orbital 
space and one looks for an unrestricted configuration with orbital order~\cite{Ave19}. 
An orbitally ordered state with breaking of the $\pi_g^x$/$\pi_g^y$ symmetry can 
only be obtained if different orbital occupations are initiated at the beginning 
of optimization~\cite{kovacik.ederer.09}, as described in Appendix~\ref{sec.appA}.   
It should be also noted that the dynamical mean-field theory (DMFT)
calculations performed for the paramagnetic Mott insulating state revealed the orbital fluctuations at high temperatures 
and the orbital order emerging only in the low-temperature (distorted) phase~\cite{kim.min.14}.
The results obtained for KO$_2$ can be compared with the DFT+DMFT studies for RbO$_2$, which
show the paramagnetic Mott insulating state at room temperature,
obtained without including dynamic distortions of the tetragonal structure, 
and indicate a complex spin-orbital order at low temperatures~\cite{kovacik.werner.12}.

Lattice distortions present in the monoclinic cell lower the symmetry of crystal 
field surrounding the superoxide anion and the degeneracy between the $\pi_g$ 
orbitals is now broken. Indeed, we find a two-sublattice orbital order in the 
density plot shown in Fig.~\ref{fig.c2_oo}(c) for the $b_{m}c_{m}$ plane, with 
alternating orbital states along the $c_m$ axis. The density distribution 
reflects the orbital states  in the electrostatic fields of K$^+$
ions after distortion, in the bases $\{\alpha',\beta'\}$ (sublattice A) and 
$\{\alpha,\beta\}$ (sublattice B), in analogy to the occupied linear combinations of
$\{t_{2g}\}$ orbitals in the vanadium perovskites with defects~\cite{Ave19}.

We obtain the same orbital order regardless whether the Coulomb $U$ is included 
(we performed calculations for $U=0$ for the unrelaxed structure). These results 
suggest the dominating role of monoclinic distortion in determining the orbital 
state. The exchange effects are too small to influence the orbital pattern, 
and we can also expect that they are highly frustrated.

A similar two-sublattice orbital order was found in an earlier study for KO$_2$~\cite{nandy.mahadevan.10}. Interestingly, within another approach~\cite{kim.kim.10,kim.min.14}, potassium ions remained in their tetragonal positions 
and coherent tilting of oxygen molecules with accompanying ferro-orbital order was 
considered.

\section{Origin of the insulating gap}
\label{sec.gap}

At first sight, the origin of the insulating gap which opens at the Fermi level seems 
to be qualitatively similar in the tetragonal and monoclinic phases. Firstly, in both 
cases the insulating gap sets inside the bands exhibiting the $\pi_g$ character. 
Secondly, the gap size is ``just'' around twice larger in the monoclinic than in 
the tetragonal phase---which can be explained by invoking the shifts of the 
potassium ions and tilting of the O$_2$ molecules in the monoclinic phase 
(cf.~Appendix.~\ref{sec.appB}). Indeed, a more detailed study of the origin 
of the insulating gap presented below confirms the similarities between the 
two cases---albeit with one, rather important, difference.

\subsection{Tetragonal phase}

Let us start with the tetragonal phase and investigate how the electronic DOS depends 
on two crucial on-site interactions present in the system: the spin-orbit coupling 
and the Hubbard-like Coulomb repulsion $U$. To this end, we calculate the DOS using 
GGA+SOC, GGA+$U$, as well as GGA+$U$+SOC approaches, cf. Fig.~\ref{fig.edos} left 
panels from top to bottom, respectively. We note in passing that the unshown ``pure'' 
GGA result cannot be distinguished from the GGA+SOC case [Fig.~\ref{fig.edos}(a)].  
Whereas the assignment of the band character to particular molecular orbitals does 
not change upon including the on-site terms in the calculations, the insulating gap 
{\it only} opens in the GGA+$U$+SOC case [Fig.~\ref{fig.edos}(e)], i.e., when 
{\it both} types of the on-site interactions are included. The explanation of this 
phenomenon is three-fold: 

First, we note that when both interactions are excluded, as in the GGA approach, 
the $\pi_g$ orbitals around the Fermi levels are degenerate 
(see~Sec. \ref{sec.orbital}). 
Second, even when the spin-orbit coupling is included, its value is too small to open 
the gap (the band width is much bigger than the band splitting induced by the SOC).
Third, on the mean-field level that is captured by the DFT+$U$ calculations, 
the Hubbard $U$ can only induce the insulating gap when the existence of the 
non-degenerate $\pi_g$ bands is allowed by the spin-orbit coupling [Fig.~\ref{fig.edos}(e)]. 
We stress that the latter leads to the distinct energies of the 
$\pi^x_g\pm i\pi^y_g$ orbitals~\cite{solovyev.08}, i.e., orbitals whose charge 
densities are exactly the same.

There is however one caveat: 
As already mentioned above, in calculations with the tetragonal symmetry, breaking 
of the orbital degeneracy of the $\pi_g$ bands with distinct charge densities is 
not allowed. The orbital order can be nevertheless ``artificially'' initiated, 
leading to a stable state and the opening of the gap of $0.65$~eV 
(see Appendix~\ref{sec.appA}).

\subsection{Monoclinic phase}

Let us now move to the monoclinic structure, and following the same sequence of 
calculation schemes, examine the impact of monoclinic distortion on the electronic DOS, 
see Fig.~\ref{fig.edos} (right panels). In contrast to the tetragonal case, without $U$  
we are unable to stabilize the structure shown in Fig.~\ref{fig.mom_mag}(b) 
\footnote{For $U=0$, the deviation of O$_2^-$ molecules from the tetragonal $c_t$ axis 
as well as potassium ions displacements are substantially reduced. Therefore, in order to 
investigate the impact of the full monoclinic distortion, we use the unrelaxed geometry.},
it is interesting though to calculate the DOS with $U=0$ for the unrelaxed cell 
(i.e., the structure optimized with $U=4$~eV).
Both in the ``pure'' GGA (unshown) and in the GGA+SOC approach [Fig.~\ref{fig.edos}(b)],  
we observe a small gap of $0.2$~eV (the splitting is however much larger than the energy 
scale of the SOC). This result, together with the orbital order reported in 
Sec.~\ref{sec.orbital}, is an indication of the presence of the Jahn-Teller effect 
in KO$_2$, in agreement with Nandy et al.~\cite{nandy.mahadevan.10}. 

\begin{figure}[!t]
\includegraphics[width=\linewidth]{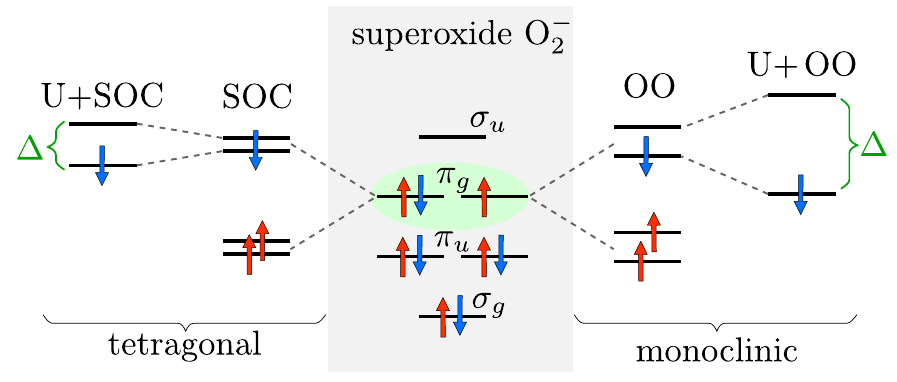}
\caption{
Schematic view of the mechanism contributing to the opening of insulating 
(Mott-Hubbard) gap in KO$_2$ by splitting degenerate $\pi_{g}$ 
orbitals located at the Fermi level in the tetragonal (left) and monoclinic 
(right) phase. In both structures the energy levels of spins up and down are 
separated by spin polarization gap. The spin-orbit coupling (SOC) leads to lifting 
of the orbital degeneracy in the tetragonal phase. The intrinsic orbital order 
(OO) induced by the Jahn-Teller effect plays a similar role in the monoclinic 
structure. In both phases, finite interaction $U$ plays a crucial role and 
enhances the insulating gap to its finite value $\Delta$. }
\label{fig.gaps}
\end{figure}

The fact that the monoclinic crystal structure leads to the splitting of the two $\pi_g$ 
orbitals around the Fermi level means that the system is already ``prepared'' for the 
$U$-induced gap and the SOC which mixes the orbital states is not required. Indeed,  
when the Hubbard $U$ is included [Fig.~\ref{fig.edos}(d)] the gap of $1.1$~eV is 
obtained, and in contrast to the tetragonal case, the impact of the SOC is negligible 
[cf. the results obtained from GGA+$U$+SOC calculations of Fig.~\ref{fig.edos}(f)]. 

The  value of the insulating gap, more than twice as large as in the tetragonal case,  
is another remarkable difference between the two phases. Whereas this phenomenon was 
discussed in a quantitative manner in Sec.~\ref{sec.electron}, it can now be naturally 
explained by invoking that the insulating gap should not only incorporate the Hubbard 
$U$-induced gap but also the `Jahn-Teller gap', as indicated by the above-discussed 
GGA (or GGA+SOC) electronic band structure. The simplified model of the monoclinic 
distortion, discussed in Appendix~\ref{sec.appB} with $U=4$~eV, reveals that even 
a small deviation from the tetragonal symmetry (just enough to lift the orbital 
degeneracy) leads to a predominantly $U$-induced gap of about $0.8$~eV. We can thus 
roughly estimate that the full monoclinic distortion contributes about 20\% to the 
insulating gap shown in Fig.~\ref{fig.edos}(d), and the remaining 80\% is the impact 
of the Coulomb $U$.

We summarize the mechanisms responsible for the opening of the gap in 
Fig.~\ref{fig.gaps} as follows. Altogether, we observe that the onset of the 
insulating gap in both tetragonal and monoclinic phases is due to the Hubbard $U$. 
On the mean-field level, this can only be realized if the $\pi_g$ orbitals 
around the Fermi level are split. 
While in the tetragonal case including a finite SOC is necessary to 
trigger the gap opening, the orbital degeneracy is broken in the monoclinic phase
due to the lattice distortions.

%%%%%%%%%%%%%%%%%%%%%%%%%%%%%%%%%%%%%%%%%%%%%%%%%%%%%%%%%%%%%%%%%%

\section{Summary}
\label{sec.summary}

In this paper, we focus on two main questions:
(i) what is the mechanism of symmetry breaking, responsible for the 
transition from the tetragonal to monoclinic phase, and 
(ii) how the crystal distortion influences the electronic properties of KO$_2$.

Regarding the first problem, we performed the {\it ab initio} studies of lattice 
dynamics for the tetragonal $I4/mmm$ phase of KO$_2$, in order to unveil the role of
phonons in the structural transition. The lowest soft mode, found at the N point of 
the Brillouin zone, induces the monoclinic distortion. The obtained $C2/c$ symmetry 
and lattice parameters of the optimized structure
agree with the diffraction data. Our study demonstrates a displacive character of the 
phase transition, in which the high- and low-temperature phases are connected by 
the group-subgroup relation. In this transition, an important role is played by the 
coupling between the soft mode and electrons, which generates the splitting of 
molecular states $\pi_g$ on oxygen anions (Jahn-Teller effect). The monoclinic 
phase is stabilized by the local Coulomb interaction $U$, which allows for 
electron localization in the orbitally ordered state.  

The implication of the structural transition on the electronic properties 
was investigated by the systematic studies of the band structures and 
electronic densities of states in both phases. In the tetragonal structure, 
the insulating state can be found by including simultaneously two on-site 
interactions: spin-orbit coupling and Hubbard $U$. In this case, a finite 
spin-orbit coupling removes the degeneracy, which triggers opening of the gap of the
order of 0.5 eV. Interestingly, initiating an orbital order in the tetragonal phase 
also leads to a stable solution  and to the 
gap of similar size---but also only once the local Hubbard interactions 
$\propto U$ are included. In the monoclinic phase, the orbital order and Mott 
insulator with the gap of the order of 1~eV is {\it naturally} stabilized by the 
lattice distortion via the Jahn-Teller effect. 

In conclusion, by performing the DFT$+U$ studies of KO$_2$, we revealed the displacive
character of the structural transition from the high-temperature tetragonal to 
low-temperature monoclinic phase. 
We observe that the onset of the insulating gap in both 
tetragonal and monoclinic phase is due to the Hubbard $U$, provided that 
the orbital degeneracy at the Fermi level is removed.

\begin{acknowledgments}
We thank Krzysztof Parlinski for valuable comments and discussions.
This work was supported by the National Science Centre (NCN, Poland) 
under grants No.:
2017/25/B/ST3/02586 (O.S., D.G., A.P., M.S., and P.P.),
2016/22/E/ST3/00560 (K.W.), and 
2016/23/B/ST3/00839 (D.G., K.W., and A.M.O.). 
A.~M.~Ole\'s is grateful for the Alexander von Humboldt
Foundation Fellowship \mbox{(Humboldt-Forschungspreis).}
\end{acknowledgments}

\appendix

\section{Tetragonal structure with orbital order}
\label{sec.appA}

Here we discuss in detail the impact of the Coulomb interaction $U$ on the 
electronic structure in the tetragonal phase. In the calculations presented 
in Fig.~\ref{fig.edos} (left panels), even large values of the Coulomb 
$U$ lead to a metallic state unless the SOC is additionally included. This can be 
understood as the result of tetragonal symmetry of the crystal structure 
(and thus charge density distribution) leading to the degeneracy of $\pi_g^x$ 
and $\pi_g^y$ orbitals. In the GGA+$U$ method, the gap 
cannot appear if the localized states around the Fermi level are degenerate.

\begin{figure}[!t]
\includegraphics[width=0.8\linewidth]{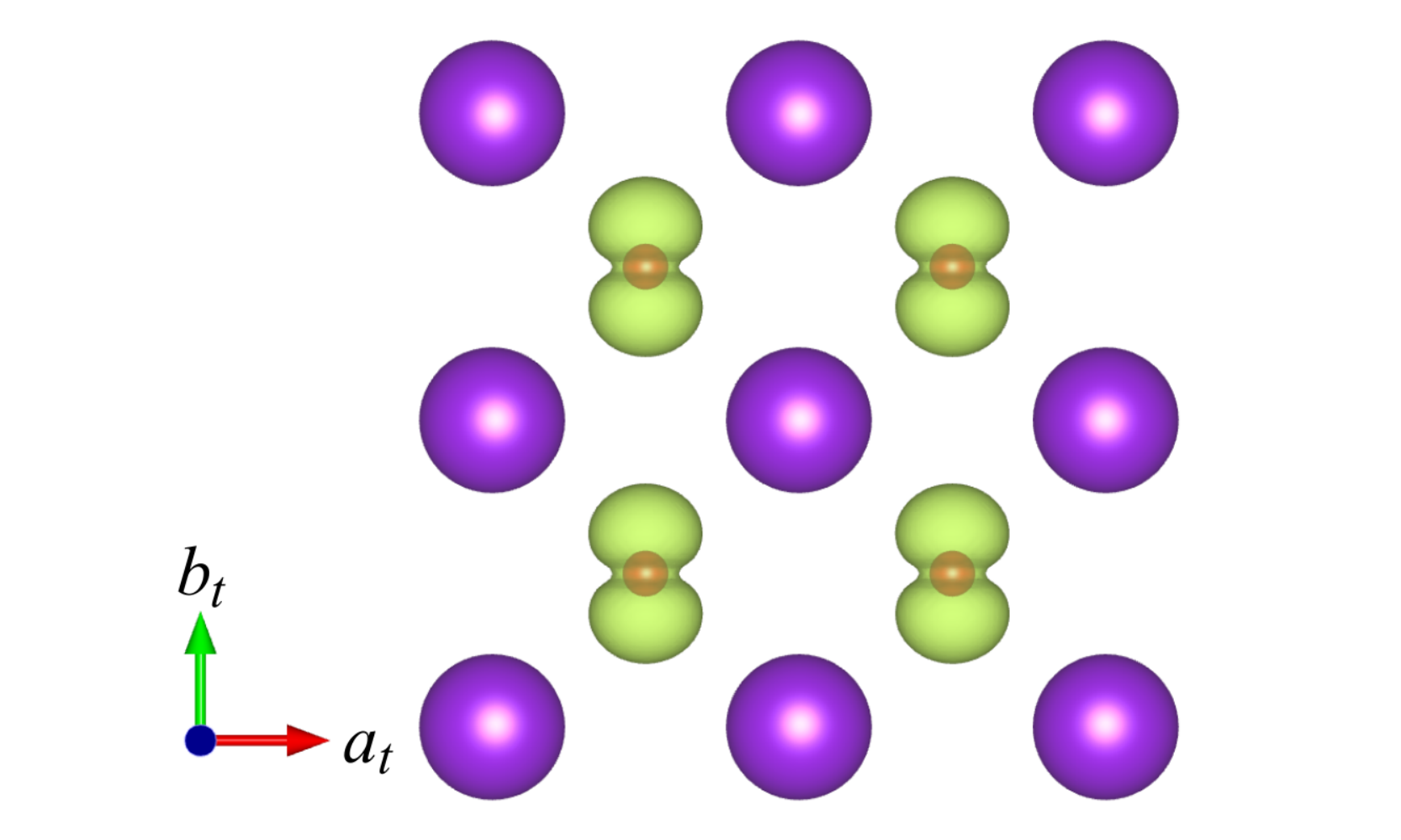}
\caption{
Calculated isosurfaces of the hole density (in green) for the tetragonal 
phase with orbital order induced by enforced symmetry breaking.
The image was rendered using {\sc vesta} software~\cite{momma.izumi.11}.
} 
\label{fig.i4_oo}
\end{figure}

\begin{figure}[!b]
\includegraphics[width=\linewidth]{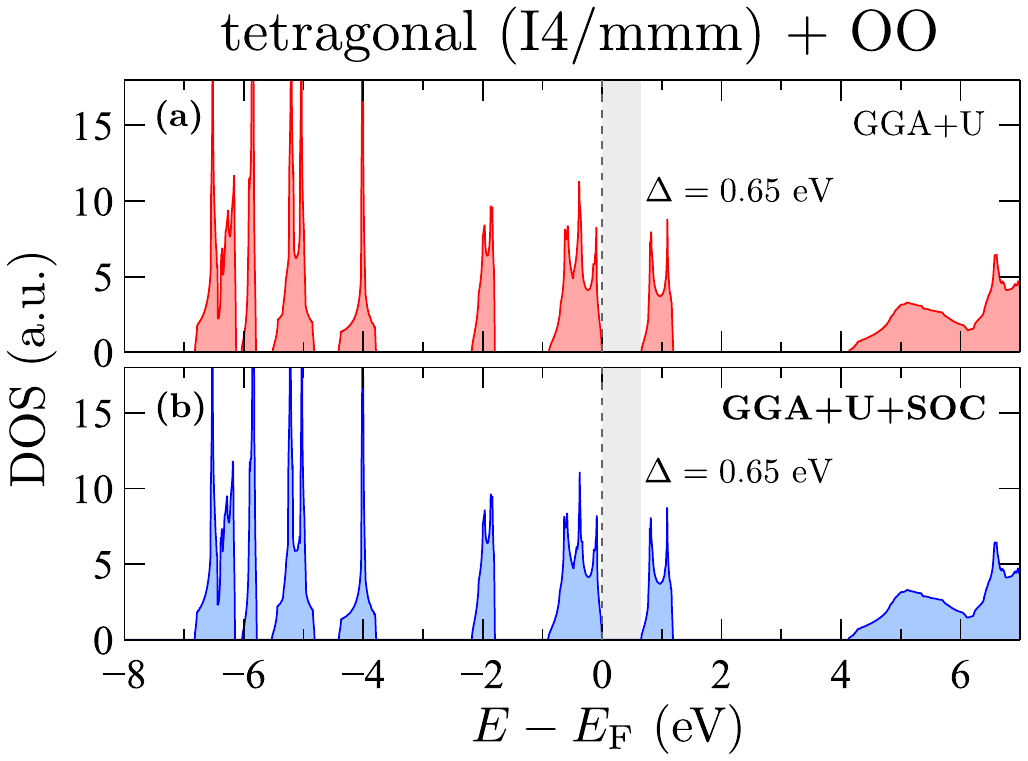}
\caption{
Electronic density of states obtained for the tetragonal phase  
with initiated orbital order (unoccupied $\pi_g^y$ orbitals) within: 
(a)~GGA+$U$ and 
(b)~GGA+$U$+SOC calculations.}
\label{fig.edos_too}
\end{figure}

\begin{figure*}[p]
\includegraphics[width=\linewidth]{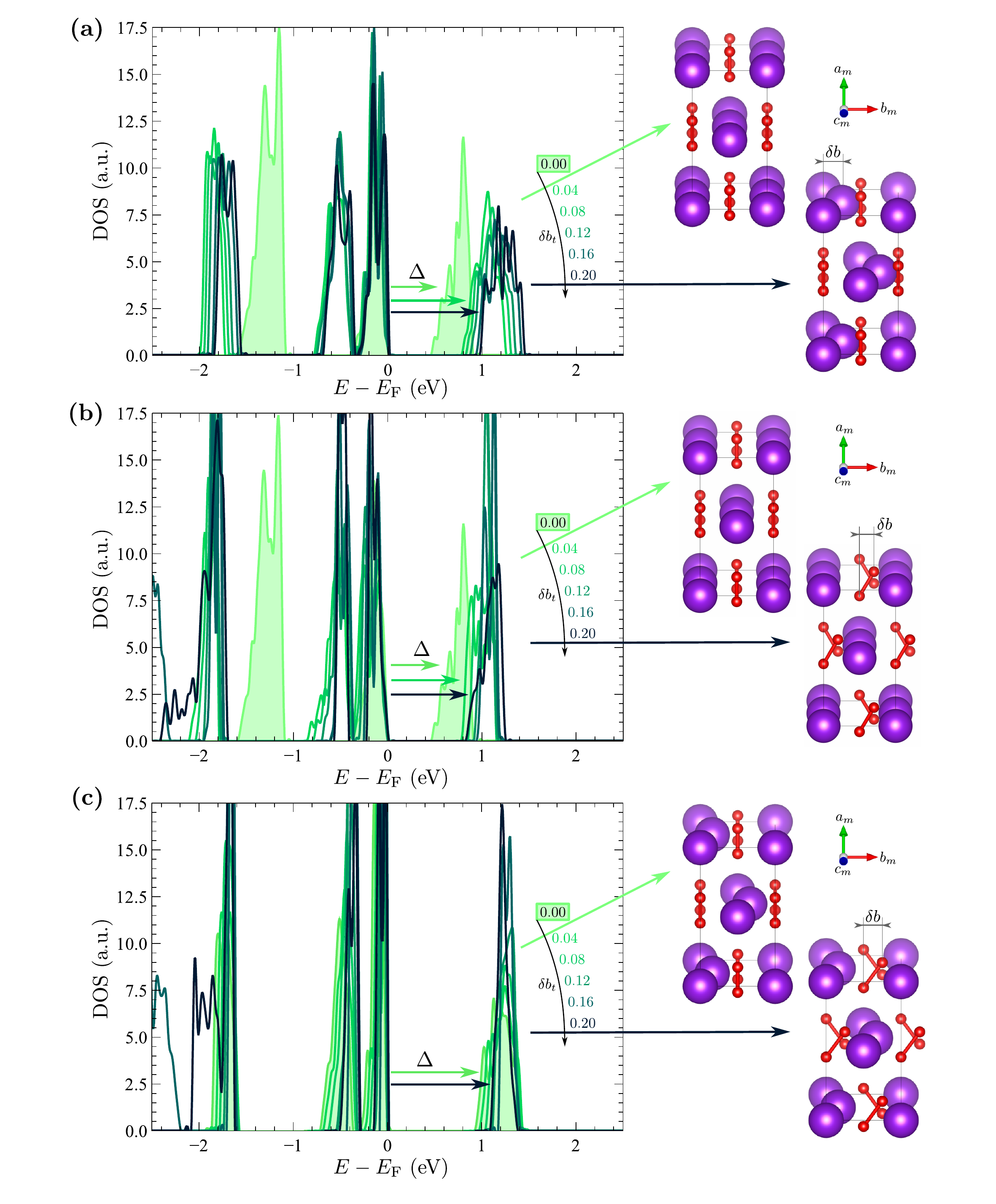}
\caption{
Comparison of the electronic density of states (DOS) for different levels of partial crystal distortions in a simplified model of the monoclinic cell: (a) positions of K atoms 
are shifted without modification of O$_{2}$ molecules; (b) oxygen molecules are tilted with K atoms in their tetragonal positions; (c) oxygen  molecules are tilted in a lattice with potassium distortions already present.
The image was rendered using {\sc vesta} software~\cite{momma.izumi.11}.
}
\label{fig.shift_dos}
\end{figure*}

It is possible however to obtain an insulating state by starting calculations from 
an orbitally ordered state and without symmetry constraint imposed on the wave 
function~\cite{kovacik.ederer.09}. We can 
demonstrate this mechanism by first calculating the charge density for a cell with 
a small orthorhombic distortion leading to the ferro-orbital ($\pi_g^y$) order shown in 
Fig.~\ref{fig.i4_oo}, and using it as an input for optimizing the tetragonal 
structure (without allowing the cell shape to change). Our results reveal that 
the orbitally ordered state has significantly lower energy ($0.2$~eV/f.u.) than 
the state optimized without an orbital order, and a gap of $0.65$~eV opens in the 
electronic density of states as shown in  Fig.~\ref{fig.edos_too}(a). Including 
the SOC [Fig.~\ref{fig.edos_too}(b)] does not induce any apparent changes to the electronic density 
of states, and this confirms that its role in opening of the gap presented 
in Fig.~\ref{fig.edos}(e) is to split the states at the Fermi level. If the orbital order is 
present in the system, including the SOC does 
not influence significantly neither the electronic structure nor the total energy 
(similarly as in the monoclinic phase with the orbital order resulting from distortion).

Calculations for several different orbital orders in 
RbO$_2$~\cite{kovacik.ederer.09} show that the lowering of total energy is similar 
in all cases (the  relative differences are of the order of $10$~meV). We note that a 
full relaxation of the system with orbital order breaking the tetragonal symmetry 
would always lead to a structural distortion.

\section{Impact of atomic displacement}
\label{sec.appB}

It is instructive to investigate the impact of atomic displacement on the insulating 
gap by evaluating the density of states (DOS) for a toy model of the monoclinic 
structure, in which we introduce separately artificial modifications of positions of 
O or K atoms, varying the size of the displacements. For this study we use a double 
tetragonal cell with lattice vectors chosen as in the monoclinic structure, and the 
simplified ``monoclinic-like'' modifications are illustrated in 
Fig.~\ref{fig.shift_dos}. This ``toy-model'' calculations, do not include e.g. 
different lattice vectors or volume of the true monoclinc cell. 

Figure~\ref{fig.shift_dos}(a) demonstrates changes in the DOS upon shifting the 
potassium ions along the $b_{m}$ axis by distance $\delta b$, where $\delta b$ 
changes from $0.0$ to $0.2$ of $b_{m}$ (the last value roughly corresponds to 
the extent of the monoclinic distortions). The results shown in light green with 
green background correspond to the tetragonal phase within the GGA+$U$+SOC scheme
[shown in Fig.~\ref{fig.edos}(e) in the main part], with a small gap of about $0.5$~eV.
We observe that the smallest displacement with $\delta b=0.04b_m$ causes already 
substantial changes: there is a jump in the gap size by about $0.35$~eV and a 
single peak below $-1$~eV separates into two. This can be understood by the 
different mechanism of the gap opening described in this work: 
even a small distortion from the tetragonal symmetry removes the degeneracy between 
 $\pi_g$ orbitals, and the DOS is primarily determined by the cooperative effect 
of the resulting orbital order and the Coulomb $U$. The impact of the smallest 
$\delta b$ shown here provides another illustration of this mechanism. Further 
displacement modifies the electronic DOS only quantitatively, and the gap 
increases gradually to about $1$~eV.

Figure~\ref{fig.shift_dos}(b) shows the analogue results for oxygen molecule 
rotations (with potassium ions in the tetragonal positions). Again, we observe 
a jump in the gap size after the smallest rotation, and then only minor changes 
(the gap varies between $0.8$~eV and $1$~eV).

The last panel, Fig.~\ref{fig.shift_dos}(c), presents modifications of the 
electronic DOS upon oxygen molecule rotations, but starting from the structure 
with shifted potassium positions. As the orbital order and considerable distortions 
are present from the beginning, we observe only quantitative modifications of the DOS. 
It is worth noting, however, that only by simultaneous displacements of the potassium 
ions and rotations of oxygen molecules, we are able to obtain the gap of about $1.1$~eV, 
i.e., similar to the result for the fully optimized monoclinic structure.

\bibliography{biblio.bib}

\end{document}